\documentclass[10pt,aps,prd,amsmath,amssymb,twoside,showpacs,nofootinbib,preprintnumbers]{revtex4}
\usepackage[T1]{fontenc}
\usepackage{amsmath,amsfonts,amssymb,amsthm,esint,wasysym}
\usepackage{mathrsfs} 
\usepackage{enumerate}
\usepackage[matrix,arrow]{xy} 
\usepackage{pictexwd}

\newcommand{\defeq}{:=}
\newcommand{\im}{\mathrm{i}}  

\newcommand{\R}{\mathbb{R}}
\newcommand{\C}{\mathbb{C}}
\newcommand{\xd}{\mathrm{d}}
\newcommand{\xD}{\mathcal{D}} 

\newcommand{\cH}{\mathcal{H}}

\newcommand{\be}{\begin{equation}}
\newcommand{\ee}{\end{equation}}
\newcommand{\bea}{\begin{eqnarray}}
\newcommand{\eea}{\end{eqnarray}}

\begin{document}

\title{Quantum field theory on timelike hypersurfaces in Rindler space}
\author{Daniele Colosi}\email{colosi@matmor.unam.mx}\affiliation{Centro de Ciencias Matem\'aticas,\\ Universidad Nacional Aut\'onoma de M\'exico, Campus Morelia, C.P.~58190, Morelia, Michoac\'an, Mexico}
\author{Dennis R\"atzel}\email{dennis.raetzel@aei.mpg.de}\affiliation{Albert Einstein Institute, Max Planck Institute for Gravitational Physics,\\ Am M\"uhlenberg 1, 14476 Golm, Germany}

\date{\today}
\preprint{UNAM-CCM-2013-1}

\begin{abstract}
The general boundary formulation of quantum field theory is applied to a massive scalar field in two dimensional Rindler space. The field is quantized according to both the Schr\"odinger-Feynman quantization prescription and the holomorphic one in two different spacetime regions: a region bounded by two Cauchy surfaces and a region bounded by one timelike curve. An isomorphism is constructed between the Hilbert spaces associated with these two boundaries. This isomorphism preserves the probabilities that can be extracted from the free and the interacting quantum field theories, proving the equivalence of the S-matrices defined in the two settings, when both apply.
\end{abstract}

\maketitle

\section{Introduction}

The general boundary formulation (GBF) provides a new axiomatic approach to describe the dynamics of quantum fields \cite{Oe:GBQFT,Oeckl2005,Oe:2dqy,Oe:hol,Oe:Sch-hol,Oe:SFobs,Oeckl2012b,Oe:aff,Oe:Obs,CoDoOe:AdS,CoDo:gen,Co:vac,CoOe:unit,CoOe:2d,CoOe:letter,CoOe:Smatrix,Oeckl:2012ni}. The set of axioms, inspired by topological quantum field theory \cite{Atiyah1989a,Oeckl:2003vu}, assigns algebraic structures to geometrical ones and ensures the consistency of these assignments. In particular, amplitude maps are associated with general spacetime regions and state spaces with their corresponding boundaries. A generalization of the Born's rule \cite{Oe:probgbf} guaranties a consistent physical interpretation of such structures.

The main motivation for the development of the GBF has been represented by conceptual difficulties inherent in the attempt to formulate a quantum theory of gravity \cite{Oeckl:2003vu,Co:GBF-QG} like the so called problem of time \cite{Isham1992}, the problem of providing a fully local description of the quantum dynamics in a quantum gravitational context and the measurement problem. From this perspective, a remarkable aspect of the GBF is the following: no background metric is required for the implementation of the GBF. 

On the one hand, it is very useful to consider quantum field theories of matter fields on fixed Lorentzian spacetimes to test the GBF and to gain insight into its structure. On the other hand, in the standard formulation of these field theories, only regions with spacelike initial and final data hypersurfaces are usually considered. Within the GBF a much wider class of setups can be implemented. Indeed, the GBF offers the possibility to construct QFTs in general spacetime regions, in particular compact spacetime regions with just one connected boundary with spacelike and timelike parts. This means that the GBF enables us to have a completely new perspective on the well-established quantum theory of matter fields.

In recent years, the GBF was applied to many different physical setups \cite{Oe:2dqy,CoDoOe:AdS,CoDo:gen,Co:vac,CoOe:unit,CoOe:2d,CoOe:letter,CoOe:Smatrix,Colosi2012}
which led to many interesting results, like the crossing symmetry of the S-matrix of perturbative quantum field theory which is a derived property within the GBF \cite{CoOe:letter,CoOe:Smatrix} or the rigorous construction in Anti-de Sitter space \cite{CoDoOe:AdS} of an asymptotic amplitude that can be interpreted as an S-matrix for \textit{spatial asymptotic states}.

In this article, we apply the GBF to study the quantum theory of a massive scalar field in 2d Rindler space in two different spacetime regions: a region bounded by two Cauchy surfaces given by hyperplanes of constant Rindler time, and a region bounded by one timelike hypersurface of constant Rindler spatial coordinate. The first region is usually considered in the standard formulation of QFT and represents an important test for the ability of the GBF to reproduce known results. In contrast, the timelike boundary of the second region makes the applicability of the standard techniques of quantization difficult and represents a significant departure from the traditional description of dynamics in QFT. We will show that the GBF can deal with this second setting with no difficulty and moreover we proof that a one-to-one relation can be established between the state spaces in the two settings. This result extends previous results obtained in Minkowski space \cite{CoOe:letter,CoOe:Smatrix}\footnote{The one-to-one correspondence established for the standard spacelike bounded regions in Minkowski space and a particular family of regions with timelike boundaries was used, in particular, to show explicitly that the crossing symmetry of QFT is generic in the GBF.} and de Sitter spaces \cite{Co:letter,Co:dS}.

The article is structured as follows: In Section \ref{sec:GBF}, we introduce the GBF and its main structures. In Section \ref{sec:ct}, the two spacetime regions of interest here are introduced and the solutions of the classical equations of motion specified. In Section \ref{sec:qt}, we present the quantization of the scalar field in both regions and in Section \ref{sec:identification}, we establish an isomorphism between the two quantum theories and show that it preserves amplitudes and probabilities in the free quantum field theory. In Section \ref{app:Feyn}, we show that this is also true for the interacting theory. Our conclusions and outlooks are summarized in Section \ref{sec:conclusions}.


\section{The General Boundary Formulation of Quantum Field Theory}
\label{sec:GBF}

In this section, we give a short review on the Schr\"odinger-Feynman representation \cite{Oe:GBQFT} and the holomorphic representation \cite{Oe:hol} in which the GBF axioms presented in \cite{Oe:GBQFT} have been so far implemented. We introduce the main structures that will be used in the rest of the paper such as state spaces and amplitude maps for both representations.

Let $S_M(\phi)=\int_M \xd^Nx\,\mathcal{L}(\phi,\partial\phi,x)$ be the action of a linear real scalar field theory in a spacetime region $M$ of an $N$-dimensional Lorentzian manifold $(\mathcal{M},g)$. 
Denoting the boundary\footnote{Notice that whether the boundary hypersurface $\Sigma$ is a Cauchy surface (or a disjoint union of Cauchy surfaces) has no bearing on the following treatment.} of the region $M$ with $\Sigma$, we associate with this hypersurface the space $L_\Sigma$ of solutions of the Euler-Lagrange equations defined in a neighborhood of $\Sigma$.\footnote{More precisely $L_\Sigma$ is the space of germs of solutions at $\Sigma$ which is the set of all equivalence classes of solutions where two solutions are equivalent if there exists a neighborhood of $\Sigma$ such that the two solutions coincide in this whole neighborhood.} The symplectic potential on $\Sigma$ results to be
\begin{equation}
 (\theta_\Sigma)_\phi(X):=\int_\Sigma \xd^{N-1}\sigma\,X(x(\sigma)) \left(n^\mu \frac{\delta\mathcal{L}}{\delta \partial_\mu\phi}\right)(x(\sigma)),
\end{equation}
where $n^{\mu}$ is the unit normal vector to $\Sigma$.
For every two elements of the space $L_\Sigma$ there is the bilinear map $[\cdot,\cdot]_{\Sigma} : L_{\Sigma} \times L_{\Sigma} \rightarrow {\mathbb R}$ defined such that $[\xi,\eta]_{\Sigma}:=(\theta_\Sigma)_\xi(\eta)$ and the symplectic structure: the anti-symmetric bilinear map $\omega_{\Sigma} :L_{\Sigma} \times L_{\Sigma} \rightarrow {\mathbb R}$ given by $\omega_{\Sigma}(\xi,\eta):=\frac{1}{2}[\xi,\eta]_{\Sigma}-\frac{1}{2}[\eta,\xi]_{\Sigma}$. The last ingredient for the quantum theory we need to specify is a compatible complex structure $J_\Sigma$ represented by the linear map $J_{\Sigma} : L_{\Sigma} \rightarrow L_{\Sigma}$ such that $J_\Sigma^2=-\text{id}$ and $\omega_{\Sigma}(J_\Sigma \cdot,J_\Sigma \cdot)=\omega_{\Sigma}(\cdot,\cdot)$ and $\omega_\Sigma(\cdot,J_\Sigma\cdot)$ is a positive definite bi-linear map. Remark, that all ingredients but the complex structure $J_\Sigma$ are classical data uniquely defined by specifying the action.

These basic ingredients can now be used in different ways to specify the Hilbert spaces which, according to the axioms of the GBF, are associated with the boundary hypersurface $\Sigma$.\footnote{If the boundary of the region considered is given by the disjoint union of two hypersurfaces, say $\Sigma = \Sigma_1 \cup \Sigma_2$, the associated Hilbert space is tensor product of the Hilbert spaces defined on each hypersurface, $\cH_{\Sigma} = \cH_{\Sigma_1} \otimes \cH_{\Sigma_2}^*$, where the different orientation of the hypersurface $\Sigma_2$ with respect to $\Sigma_1$ is responsible for the dualization of the corresponding Hilbert space.} In the following sections, we introduce the two representations developed so far within the GBF, namely the Schr\"odinger representation, usually associated with the Feynman path integral quantization prescription, and the holomorphic representation.

\subsection{The holomorphic representation}

From the complex structure $J_\Sigma$ we define the symmetric bilinear form $g_\Sigma : L_\Sigma \times L_\Sigma \rightarrow {\mathbb R}$ as
\begin{equation}
g_\Sigma(\xi,\eta):=2\omega_\Sigma(\xi,J_\Sigma\eta) \qquad \forall \xi, \eta \in L_\Sigma,
\end{equation}
and assume that this form is positive definite. Next, we introduce the sesquilinear form 
\begin{equation}
\{\xi,\eta\}_\Sigma:=g_\Sigma(\xi,\eta)+2 \im \omega_\Sigma(\xi,\eta) \qquad \forall \xi, \eta \in L_\Sigma.
\end{equation}
The completion of $L_\Sigma$ with the inner product $\{ \cdot, \cdot \}_\Sigma$ turns it into a complex Hilbert space.
The Hilbert space $\mathcal{H}^h_\Sigma=H^2(L_\Sigma,\xd\nu_{\Sigma}),$\footnote{To make this mathematically precise one actually has to construct $\mathcal{H}^h_\Sigma= H^2(\hat L_{\Sigma},\xd\nu_{\Sigma})$ where $\hat L_{\Sigma}$ is a certain extension of $L_{\Sigma}$. For more details about the construction of $\hat L_{\Sigma}$ and $\xd\nu_{\Sigma}$ we refer the reader to \cite{Oe:hol}.} namely the set of square integrable holomorphic functions on $L_\Sigma$, is the closure of the set of all coherent states\footnote{States in the holomorphic representation are denoted with a superscript $h$.} \cite{Oe:hol}
\begin{equation}
 K^h_{\Sigma,\xi}(\phi):=e^{\frac{1}{2}\{\xi,\phi\}_{\Sigma}},
\end{equation}
where $\xi \in L_\Sigma$ and the closure is taken with respect to the inner product
\begin{equation}
 \langle K^h_{\Sigma,\xi}, K^h_{\Sigma,\xi'}\rangle:=\int_{L_\Sigma} \xd \nu_\Sigma(\phi)\, \overline{K^h_{\Sigma,\xi}(\phi)}K^h_{\Sigma,\xi'}(\phi),
\end{equation}
where $\xd\nu_\Sigma$ is a Gaussian probability measure constructed from the metric $g_\Sigma$ \cite{Oe:hol}. It can be represented formally as $\xd \nu_\Sigma(\phi)=\xd\mu_\Sigma(\phi)e^{\frac{1}{4}g_\Sigma(\phi,\phi)}$ with a certain translation invariant measure $\xd\mu_\Sigma$. 

Associated to each spacetime region $M$ there is an amplitude $\varrho_M$ defined for states belonging to the Hilbert space associated with the boundary $\Sigma$ of this region, 
\begin{equation}
 \varrho_M(\psi^h):=\int_{L_{\tilde M}} \xd\nu_{\tilde M}(\phi)\,\psi^h(\phi),
\end{equation}
where $L_{\tilde{M}}\subseteq L_\Sigma$ is the set of all global solutions on $M$ mapped to $L_\Sigma$ by just considering the solutions in a neighborhood of $\Sigma$.\footnote{More precisely, global solutions are mapped to the corresponding germs at $\Sigma$.} The measure $\xd\nu_{\tilde M}$ is again a Gaussian probability measure constructed from the metric $g_\Sigma$ \cite{Oe:hol}.\footnote{Again, we refer the reader to \cite{Oe:hol} where the constructions are given that make all the objects used here well defined. Additionally, in \cite{Oe:Sch-hol} it was shown that the one-to-one correspondence between maps $\Omega_\Sigma$, which is an important ingredient of the Schr\"odinger-Feynman representation and will be defined in the next section, and complex structures $J_\Sigma$ leads also to mathematically well defined constructions for all the expressions in section \ref{sec:Schroedinger}.}
This amplitude for coherent states turns out to be\footnote{See equation (31) of \cite{Oe:SFobs} for normalized coherent states and equation (43) in \cite{Oe:hol} as well as \cite{Oe:Sch-hol}.} 
\begin{equation}
 \varrho_M(K^h_\xi)=\exp\left(\frac{1}{4}g_\Sigma(\xi^R,\xi^R)-\frac{1}{4}g_\Sigma(\xi^I,\xi^I)-\frac{\im}{2}g_\Sigma(\xi^R,\xi^I)\right),
 \label{eq:freeamphol}
\end{equation}
where $\xi^R,\xi^I\in L_{\tilde M}$ and $\xi=\xi^R+J_\Sigma\xi^I$. A consistent probability interpretation can be given to this amplitude using the generalized Born's rule \cite{Oe:GBQFT,Oe:probgbf} defined in the GBF. 

\subsection{The Sch\"odinger-Feynman representation}\label{sec:Schroedinger}

In this section, we introduce the Schr\"odinger-Feynman representation of the GBF. However, we will not start from the symplectic form and complex structure we established in the beginning but directly from the action $S_M(\phi)$. This is the way the Schr\"odinger-Feynman representation was established originally. The construction of the Schr\"odinger-Feynman representation from the symplectic form and complex structure will be the content of the next section which will illuminate the relation between the two representations.

In the Schr\"odinger-Feynman representation, quantum states in the Hilbert space associated with the boundary $\Sigma$ are represented as wave functionals of the space of field configurations\footnote{We denote states in the Schr\"odinger-Feynman representation with a superscript $S$.}. 
The amplitude associated with the region $M$ is given by the linear map $\varrho_M: \cH_{\Sigma} \rightarrow \C$,
\begin{equation}
\varrho_M(\psi^S) = \int \xD \varphi \, \psi^S(\varphi) Z_M(\varphi),
\label{eq:amplitude}
\end{equation}
where the integral is extended over all the configurations $\varphi$ on the boundary of the region $M$, 
and $Z_M(\varphi)$ is the field propagator, formally defined as
\begin{equation}
Z_M(\varphi) = \int_{\phi|_{\Sigma}=\varphi} \xD\phi\, e^{\im S_{M}(\phi)},
\label{eq:proppint0}
\end{equation}
where $S_M(\phi)$ is the action of the field in $M$ and the integral is extended to the spacetime field configurations $\phi$ that reduce to the configuration $\varphi$ on the boundary hypersurface $\Sigma$.

As in the holomorphic representation, coherent states can be defined in the Schr\"odinger representation. They are given as
\begin{equation}
K^S_{\Sigma, \xi}(\varphi) = \kappa_{\Sigma, \xi} \, \exp \left( \int \xd ^3 s \, \xi(s) \varphi(s) - \frac{1}{2} \Omega_{\Sigma}(\varphi,\varphi) \right),
\label{eq:csSF}
\end{equation}
where $\kappa_{\Sigma,\xi}$ is a normalization constant and $\Omega_{\Sigma}$ is a bilinear map from two copies of the space of field configurations on the boundary hypersurface $\Sigma$ to the complex numbers. The vacuum state is obtained from (\ref{eq:csSF}) by setting $\xi=0$.

With the coherent states above we can again define the Hilbert space associated with the boundary $\Sigma$ as the closure of the space of coherent states with respect to an inner product. In the Schr\"odinger representation this is the expression
\begin{equation}
  \langle \psi_{\Sigma} | \psi_{\Sigma}' \rangle \defeq \int \xD \varphi \, \overline{\psi_\Sigma(\varphi)} \, \psi_{\Sigma}'(\varphi).
\end{equation}
                                                                                                                                                         \subsection{Relation between the two representations}

In this section we show how to develop the Schr\"odinger-Feynman representation starting from the symplectic form and the complex structure. We also clarify the relation between the two representations.

We start by defining what plays the role of the "space of momentum" in the Schr\"odinger-Feynman representation:
\be
M_\Sigma:=\{ \eta\in L_\Sigma:[\xi,\eta]=0\,\forall\xi\in L_\Sigma  \}.
\ee 
It can be shown that $M_\Sigma$ is a Lagrangian subspace of $L_\Sigma$.\footnote{It is this subspace $M_\Sigma$ that defines the Schr\"odinger polarization of the prequantum Hilbert space constructed from $L_\Sigma$, see \cite{Oe:Sch-hol} for details.} Next, we consider the quotient space 
$Q_\Sigma:=L_\Sigma/M_\Sigma$ which corresponds to the space of all field configurations on $\Sigma$. We denote the quotient map $L_\Sigma \rightarrow Q_\Sigma$ by $q_\Sigma$. The last ingredient needed for the Schr\"odinger representation is the bilinear map defining the vacuum state,
\begin{align}
\Omega_\Sigma: &Q_\Sigma \times Q_\Sigma \rightarrow \mathbb{C}, \nonumber\\
&(\varphi,\varphi')\mapsto 2\omega_\Sigma(j_\Sigma(\varphi),J_\Sigma j_\Sigma(\varphi'))- \im [j_\Sigma(\varphi),\varphi']_\Sigma,
\end{align} 
where $j_\Sigma$ is the unique linear map $Q_\Sigma \rightarrow L_\Sigma$ such that $q_\Sigma \circ j_\Sigma = \text{id}_{Q_\Sigma}$ and $j_\Sigma (Q_\Sigma) \subseteq J_\Sigma M$.
Coherent states are given in terms of $\Omega_\Sigma$ by the expressions
\begin{equation}
  K^S_{\Sigma,\xi}(\varphi)=\exp\left(\Omega_\Sigma(q_\Sigma(\xi),\varphi)+ \im [\xi,\varphi]_\Sigma-\frac{1}{2}\Omega_\Sigma(q_\Sigma(\xi),q_\Sigma(\xi))-\frac{\im}{2}[\xi,\xi]_\Sigma-\frac{1}{2}\Omega_\Sigma(\varphi,\varphi))\right).
\label{eq:csSF-J}
\end{equation} 
It was shown in \cite{Oe:Sch-hol} that there is a one-to-one correspondence between bilinear maps $\Omega_\Sigma$ appropriate for the Schr\"odinger representation and complex structures $J_\Sigma$. This means that given a complex structure, we uniquely fix all the algebraic structures of the two representations.\footnote{This one-to-one correspondence sends (\ref{eq:csSF-J}) into (\ref{eq:csSF}).} In particular, an isomorphism exists between the Hilbert spaces in the holomorphic representation and the Schr\"odinger-Feynman representation that preserves the amplitude map.
Hence, the two representations can be used equivalently.

\section{Classical theory}
\label{sec:ct}

Rindler space $\mathcal{R}$ is given by the metric $ds^2 = \rho^2 d \eta^2 - d \rho^2$ where $\rho\in \mathbb{R}^+$ and $\eta\in\mathbb{R}$. 
The free action of the Klein-Gordon field in a spacetime region $M$ is  
\begin{equation}
S_{M,0}(\phi)=\frac{1}{2}\int_{M} \xd \eta \, \xd \rho \, \rho \left(- \left(\partial_{\rho} \phi\right)^2 + \frac{1}{\rho^2}\left(\partial_{\eta} \phi\right)^2  - m^2\phi^2\right),
\label{eq:freeact}
\end{equation}
where $\partial_{\rho}$ and $\partial_{\eta}$ denote the partial derivatives with respect to $\rho$ and $\eta$ respectively. From the action we can deduce the equation of motion,
\begin{equation}
\left( -\rho \partial_{\rho} \rho \partial_{\rho} + \partial_{\eta}^2 + m^2 \rho^2 \right) \phi = 0.
\label{eq:eom}
\end{equation}
Solutions of the field equation (\ref{eq:eom}) can be expressed in terms of the modes
\be
\chi_p(x) = \frac{\im}{2}(\sinh (p \pi))^{-1/2} I_{\im p} (m \rho) e^{-\im p \eta}, \qquad \phi_p(x) = \frac{(\sinh (p \pi))^{1/2}}{\pi} K_{\im p} (m \rho) e^{-\im p \eta}, \qquad p \geq 0.
\label{eq:Kmodes}
\ee
where $I_{\im p}$ and $K_{\im p}$ are the modified Bessel functions of the first and second kind respectively, see Appendix \ref{sec:appA}.

In the following, we will study the field in two different spacetime regions: A region $M_1$ bounded by two semi-lines of constant Rindler time $\eta_1$ and $\eta_2$ respectively, with $\eta_1< \eta_2$; namely $M_1 = [\eta_1,\eta_2] \times \R^+$ and all the relevant quantities referring to this region will be indicated with the subscript $[\eta_1,\eta_2]$. Additionally, we will consider the region $M_2$ bounded by one hyperperbola of constant $\rho=\rho_1$; namely $M_2 2= \R \times [\rho_1, \infty)$. 
Because of the asymptotic behavior (\ref{eq:asympIK}), in both regions the field will be expanded in the basis of the modes $\phi_p(x)$.


\subsection{Region with spacelike boundary: $M_1$}
\label{sec:regionM1}

Consider the region bounded by the two semi-lines of constant $\eta$, namely the region $M_1$. We denote by $\varphi_1$ and $\varphi_2$ the configurations of the field on the boundaries $\Sigma_1$ at $\eta=\eta_1$ and $\Sigma_2$ at $\eta= \eta_2$, respectively: $\phi|_{\Sigma_1}=\varphi_1$ and $\phi|_{\Sigma_2}=\varphi_2$. It will be useful to express the solution of the Klein-Gordon equation (\ref{eq:eom}) in terms of these boundary field configurations. In particular, the general solution to equation (\ref{eq:eom}) can be written as
\begin{equation}
\phi(\eta, \rho) = \left(X_a(\eta) Y_a \right)(\rho) + \left(X_b(\eta) Y_b \right)(\rho),
\end{equation}
where each $X_i(\eta)$ is understood as an operator acting on a mode decomposition of $Y_i$. In particular we can choose $X_a(\eta)= \cos(p \eta)$ and $X_b(\eta)= \sin (p \eta)$. Expressing each $Y_i$ in terms of the boundary field configurations $\varphi_i$ leads to expression, 
\begin{equation}
\phi(\eta, \rho) = \left( \frac{\sin p (\eta_2-\eta)}{\sin p (\eta_2-\eta_1)} \varphi_1 \right) (\rho) + \left( \frac{\sin p (\eta-\eta_1)}{\sin p (\eta_2-\eta_1)} \varphi_2 \right) (\rho),
\label{eq:phiM1}
\end{equation}
where $p$ is to be understood as the operator $p \defeq \sqrt{(\rho \partial_{\rho})^2-m^2}$ acting on a mode decomposition of the boundary field configurations. As mentioned above, the divergent character of $I_{\im p}$ at infinity forces us to retain in this mode expansion only the modified Bessel function of the second kind, $K_{\im p}$, also known as Macdonald function, see Appendix \ref{sec:appA}.
The free action (\ref{eq:freeact}) in terms of the boundary field configurations reads
\begin{equation}
S_{[\eta_1,\eta_2],0}(\varphi_1,\varphi_2) =  \frac{1}{2} \int_0^{\infty} \frac{\xd \rho}{\rho} \, 
\begin{pmatrix}\varphi_1 & \varphi_2 \end{pmatrix} W_{[\eta_1,\eta_2]}  
\begin{pmatrix} \varphi_1 \\ \varphi_2 \end{pmatrix},
\label{eq:freeactM1}
\end{equation}
where the $W_{[\eta_1,\eta_2]}$ is a $2 \times 2$ matrix given by
\begin{equation}
W_{[\eta_1,\eta_2]}= \frac{p}{\sin p (\eta_2-\eta_1)}
\begin{pmatrix} \cos p (\eta_2-\eta_1) & -1 \\
-1 & \cos p (\eta_2-\eta_1) \end{pmatrix}.
\end{equation}


\subsection{Region with timelike boundary: $M_2$}
\label{sec:regionM3}

In contrast to the spacetime region considered before, the region $M_2$ presents only one boundary $\Sigma_{\rho_1}$ defined by the hyperbola $\rho=\rho_1$, i.e., $M_2 = \R \times [\rho_1, \infty)$. The subscript $\rho_1$ will be used for the quantities referring to this region.
The field configurations will then contain only the modified Bessel function of the first kind and a solution of the Klein-Gordon equation in this region, reducing to the boundary configuration $\varphi$ at $\rho_1$, can be written as
\begin{equation}
\phi(\eta,\rho) = \left( \frac{K_{\im p}(m \rho)}{ K_{\im p}(m \rho_1)} \, \varphi \right)(\eta)\,,
\label{eq:phiM3}
\end{equation}

where $\frac{K_{\im p}(m \rho)}{ K_{\im p}(m \rho_1)}$ has to be understood as an operator acting on the field configuration $\varphi(\eta)$ as 
\begin{equation}
 \frac{K_{\im p}(m \rho)}{ K_{\im p}(m \rho_1)}e^{ip'\eta}=\frac{K_{\im p'}(m \rho)}{ K_{\im p'}(m \rho_1)}e^{ip'\eta}\,.
\end{equation}
The action of the field (\ref{eq:phiM3}) in the region $M_2$ is expressed in terms of $\varphi$ as
\begin{equation}
S_{\rho_1,0}(\varphi) = \frac{1}{2} \int_{-\infty}^{\infty} \xd \eta \left.  \, \varphi(\eta)\, \rho\frac{\xd}{\xd \rho}\left(\frac{K_{\im p}(m \rho)}{ K_{\im p}(m \rho_1)} \, \varphi \right)(\eta)\right|_{\rho=\rho_1}\,.
\label{eq:freeactM3}
\end{equation}


\section{Quantum theory}
\label{sec:qt}
In this section, the quantum theory of the free field in the different regions considered above will be presented. In \cite{CoDo:gen}, a general treatment of the GBF description of the quantum dynamics of a scalar field in a certain class of spacetimes and spacetime regions has been presented. The scalar field in the two spacetime regions in Rindler spacetime considered here satisfies the conditions of \cite{CoDo:gen} and the results obtained can then be used in the present work.

\subsection{Quantization in $M_1$}
\label{sec:Rin}

\subsubsection{Holomorphic representation}
To constitute valid initial data on the hypersurfaces $\Sigma_1$ and $\Sigma_2$, the field $\phi$ must vanish at spacelike infinity which excludes the modes containing the Bessel functions of the first kind and leaves us with the decomposition
\begin{equation}\label{eq:phiR}
\phi (x)= \int_{0}^{\infty} \xd p \left(\phi(p)\phi_p(x) + c.c.\right).
\end{equation}
From the second variation of the action in equation (\ref{eq:freeact}), we obtain the symplectic form as
\begin{equation}
 \omega_{\Sigma_i}(\phi,\phi')=\frac{1}{2}\int_0^\infty \frac{\xd \rho}{\rho}\, \left(  \phi\partial_\eta \phi'  -\phi'\partial_\eta \phi\right)(\rho).
\end{equation}
Now, we obtain for two modes $\phi_p$ and $\phi_{p'}$ at $\Sigma_i$ with $i=1,2$ the following expressions:
\begin{equation}
\omega_{\Sigma_i}(\overline{\phi_p},\phi_{p'})= \delta(p-p'), \qquad \omega_{\Sigma_i}(\phi_p,\phi_{p'})=\omega_{\Sigma_i}(\overline{\phi_p},\overline{\phi_{p'}})=0.
\end{equation}
With the complex structure
\begin{equation}
J_{\Sigma_i}= \frac{\partial_{\eta}}{\sqrt{-\partial_{\eta}^2}},
\label{eq:cstructureRindler}
\end{equation}
which corresponds to the timelike Killing vector field $\partial_\eta$, we obtain, for two general solutions $\phi$ and $\psi$ 
\begin{align}
\omega_{\Sigma_{i}}(\phi,\psi)&=\frac{i}{2}\int_{0}^{\infty} \xd p\,\left( \overline{\phi(p)} \psi(p) - \phi(p) \overline{\psi(p)} \right), \label{eq:symstrRin} \\
g_{\Sigma_{i}}(\phi,\psi) &=\int_{0}^{\infty} \xd p\,\left( \overline{\phi(p)} \psi(p) + \phi(p) \overline{\psi(p)} \right), \label{eq:metRin} \\
\left\{ \phi,\psi\right\}_{\Sigma_{i}} &= g_{\Sigma_{i}}(\phi,\psi)+2i\omega_{\Sigma_{i}}(\phi,\psi)=2\int_{0}^{\infty} \xd p\, \phi(p) \overline{\psi(p)}. \label{eq:innproRin}
\end{align}
These are all the algebraic objects necessary for the holomorphic quantization of the Klein-Gordon field in the region $M_1$. 

\subsubsection{Schr\"odinger-Feynman representation}
Substituting in (\ref{eq:proppint0}) the free action (\ref{eq:freeactM1}) of the classical solution (\ref{eq:phiM1}) in the spacetime region $M_1$ we can express the field propagator in terms of the boundary field configurations $\varphi_1$ and $\varphi_2$,
\begin{equation}
Z_{[\eta_1,\eta_2],0}(\varphi_1,\varphi_2) = \left( \det \frac{- \im p}{2 \pi \sin p (\eta_2-\eta_1)} \right)^{-1/2} e^{\im S_{[\eta_1,\eta_2],0}(\varphi_1,\varphi_2)},
\label{eq:fpropM1}
\end{equation}
where again $p =\sqrt{(\rho \partial_{\rho})^2-m^2}$ has to be understood as an operator.
This field propagator satisfies the composition property
\begin{equation}
Z_{[\eta_1,\eta_3],0}(\varphi_1,\varphi_3) = \int \xD \varphi_2 \, Z_{[\eta_1,\eta_2],0}(\varphi_1,\varphi_2) \, Z_{[\eta_2,\eta_3],0}(\varphi_2,\varphi_3).
\label{eq:comprel1}
\end{equation}
Following \cite{CoOe:Smatrix,Co:dS,CoDo:gen} we define by (\ref{eq:csSF}) the coherent states in the Hilbert space $\cH_{\eta}$ associated to the semi line of constant Rindler time $\eta$. These states have the property to remain coherent under the evolution implemented by the field propagator (\ref{eq:fpropM1}). In the interaction picture, they take the form
\begin{align}
K^S_{\eta,\xi}(\varphi) =
\exp \left(- \frac{1}{2} \int_0^{\infty} \xd p \, \frac{1}{2 p} \left( e^{-2 \im p \eta} \xi^2(p) + |\xi(p)|^2 \right) \right)
\exp \left(\int_0^{\infty} \xd p \, e^{-\im p \eta} \xi(p) \, \varphi(p) \right) \psi_{\eta,0}(\varphi),
\label{eq:csM1}
\end{align}
where $\psi_{\eta,0}$ is the vacuum state\footnote{In the notation of \cite{CoDo:gen} the coefficients $c_a$ and $c_b$ have be chosen to be 1 and $\im$ respectively.} in $\cH_{\eta}$,
\begin{equation}
\psi_{\eta,0}(\varphi) = \det \left( \frac{p}{\pi e^{\im p \eta}}\right)^{1/4} \exp \left( -\frac{1}{2} \int  \xd p' \, \varphi(p') p' \varphi(p')\right)\,.
\label{eq:vacuum1}
\end{equation}
We have now at our disposal all the ingredients to compute explicitly the free amplitude for a coherent state in the spacetime region $M_1$. In particular we consider the coherent state defined by two complex functions $\xi_1$ and $\xi_2$ as $K^S_{\eta_1,\xi_1} \otimes \overline{K^S_{\eta_2,\xi_2}}$ in the Hilbert space $\cH_{\eta_1} \otimes \cH_{\eta_2}^*$ associated with the boundary of $M_1$. The free amplitude results to be
\begin{align}
\nonumber \varrho_{[\eta_1,\eta_2]} \left( K^S_{\eta_1,\xi_1} \otimes \overline{K^S_{\eta_2,\xi_2}} \right) &=
\int \xD \varphi_1 \, \xD \varphi_2 \, \overline{K^S_{\eta_2,\xi_2}(\varphi_2)} \, K^S_{\eta_1,\xi_1}(\varphi_1) \, Z_{[\eta_1,\eta_2],0}(\varphi_1,\varphi_2), \nonumber\\
&= \exp \left( - \frac{1}{2} \int_0^{\infty}  \frac{\xd p}{2 p }  \left(|\xi_1(p)|^2  + |\xi_2(p)|^2  - 2 \overline{\xi_2(p)} \, \xi_1(p)\right) \right),
\label{eq:freeamplM1}
\end{align}
where we used again the expansion of the functions $\xi_{1,2}(\rho)$ in the basis of the modes $u_{p}(\rho)$. Notice that this amplitude does not depend on the Rindler times $\eta_1$ and $\eta_2$. 


\subsection{Quantization in $M_2$}
\label{sec:Mh}

In this section we will give all elements of the two representations of the GBF in the region $M_2$. 

\subsubsection{Holomorphic quantization}
For the holomorphic representation, we start with the symplectic form:
\begin{equation}
\omega_{\Sigma_{\rho_1}}(\phi,\phi') = \frac{1}{2}\int_{-\infty}^\infty \xd \eta \, \rho  \left(\phi\, \partial_\rho \phi'-\phi' \partial_\rho \phi\right)(\eta).
\end{equation}
The solutions to the Klein-Gordon equation in Rindler space at $\Sigma_{\rho_1}$ only exist locally around $\Sigma_{\rho_1}$, and thus are not expected to vanish at spacelike infinity. They can be parametrized using the modified Bessel functions of the first kind as\footnote{The only solutions we have to consider at the boundary are the Bessel functions of the first kind since the Bessel functions of the second kind are not independent solutions (see (\ref{eq:besselrel})).}
\begin{equation}\label{eq:boundparam}
 \phi_{\Sigma_{\rho_1}}(x)=\int_{-\infty}^\infty \xd p \,\left(\phi_{\Sigma_{\rho_1}}(p)\chi_p(x) +c.c.\right),
\end{equation}
In contrast, solutions in the interior of $M_2$ must vanish for $\rho\to \infty$, and thus can be parameterized using just the modified Bessel functions of the second kind as
\begin{equation}\label{eq:intparam}
 \phi_{M_2}(x)=\int_{0}^\infty \xd p\, \left(\phi_{M_2}(p)\phi_p(x) +c.c.\right).
\end{equation}

For the parameterization in Equation (\ref{eq:boundparam}) and with the Wronskian of two Bessel functions of the first kind (see (\ref{eq:Wr})),
we find for the symplectic form the expression
\begin{equation}
\omega_{\Sigma_{\rho_1}}\left(\phi_{\Sigma_{\rho_1}},\phi'_{\Sigma_{\rho_1}}\right)=-\frac{\im}{2}\int_{-\infty}^{\infty} \xd p \left(\phi_{\Sigma_{\rho_1}}(p)\overline{\phi'_{\Sigma_{\rho_1}}(p)}-c.c.\right).
\end{equation}
To obtain the metric and the inner product on the space of solutions $L_{\Sigma_{\rho_1}}$ at $\Sigma_{\rho_1}$ we define the action of the complex structure $J_{\Sigma_{\rho_1}}$ as $J_{\Sigma_{\rho_1}}\chi_p(x)=- \im \chi_p(x)$. 
Hence, we obtain that
\begin{equation}
 J_{\Sigma_{\rho_1}}\phi_{\Sigma_{\rho_1}} (x)=- \im \int_{-\infty}^\infty \xd p \,\left(\phi_{\Sigma_{\rho_1}}(p)\chi_p(x) -c.c.\right),
\end{equation}
and the metric and the inner product result
\begin{align}
g_{\Sigma_{\rho_1}} \left(\phi_{\Sigma_{\rho_1}},\phi'_{\Sigma_{\rho_1}}\right) &= 2\omega_{\Sigma_M}\left(\phi_{\Sigma_{\rho_1}},J_{\Sigma_{\rho_1}}\phi'_{\Sigma_{\rho_1}}\right)= \int_{-\infty}^{\infty} \xd p\, \left(\phi_{\Sigma_{\rho_1}}(p)\overline{\phi'_{\Sigma_{\rho_1}}(p)}+c.c.\right), \\
\{\phi_{\Sigma_{\rho_1}},\phi'_{\Sigma_{\rho_1}}\}_{\Sigma_{\rho_1}} &= g_{\Sigma_{\rho_1}}\left(\phi_{\Sigma_{\rho_1}},\phi'_{\Sigma_{\rho_1}}\right)+2 \im \omega_{\Sigma_M} \left(\phi_{\Sigma_{\rho_1}},\phi'_{\Sigma_{\rho_1}}\right) = 2\int_{-\infty}^{\infty} \xd p\, \phi_{\Sigma_{\rho_1}}(p)\overline{\phi'_{\Sigma_{\rho_1}}(p)}.
\end{align}
By defining coherent states and their amplitudes we obtain the free quantum theory for the Klein-Gordon field in region $M_2$. In the next section we will establish the identification between states on the boundary of $M_2$ and $M_1$.

\subsubsection{Schr\"odinger-Feynman quantization}
The field propagator is expressed in terms of the action (\ref{eq:freeactM3}) as
\begin{equation}
Z_{\rho_1,0}(\varphi) = \det \left( \frac{4 \pi^2 K_{\im |p|}^2(m \rho_1)}{m \sinh (|p| \pi)} \right)^{-1/4} e^{\im S_{\rho_1,0}(\varphi)}.
\end{equation}
where the expression in the determinant is to understood as an operator acting as
\begin{equation}
 \frac{4 \pi^2 K_{\im |p|}^2(m \rho_1)}{m \sinh (|p| \pi)}e^{ip'\eta}=\frac{4 \pi^2 K_{\im |p'|}^2(m \rho_1)}{m \sinh (|p'| \pi)}e^{ip'\eta},
\end{equation}
on the Fourier expansion of field configurations.
We will consider the vacuum state
\begin{equation}
\psi_{\rho_1,0}(\varphi)= C_{\rho_1} \exp \left( -\frac{1}{2} \int \xd \eta \, \left.\varphi(\eta) \left( \im  \rho \frac{\xd}{\xd \rho} \ln \left(\overline{I_{\im |p|}(m \rho)}\right) \varphi \right)(\eta)\right|_{\rho=\rho_1} \right),
\label{eq:vacuum2-2}
\end{equation}
giving rise to the Hilbert space $\cH_{\Sigma_{\rho_1}}$. A coherent state in $\cH_{\Sigma_{\rho_1}}$, in the interaction picture, reads 
\begin{align}
K^S_{\rho_1,\xi}(\varphi) &= \kappa_{\rho_1,\xi} \exp \left( \int_{0}^{\infty} \frac{\xd p}{\overline{I_{\im |p|}(m \rho_1)} } \left[ \xi(p) \, \varphi(-p) + \xi(-p) \, \varphi(p) \right] \right) \psi_{\rho_1,0}(\varphi), 
\label{eq:cohM2-2}
\end{align}
where $\xi(p)$ and $\varphi(p)$ are the coefficients of the expansion of $\xi(\eta)$ and $\varphi(\eta)$ respectively in the basis of the plane waves $e^{\im p \eta}/\sqrt{2 \pi}$. $\kappa_{\rho_1, \xi}$ is the normalization factor given by 
\begin{align}
\kappa_{\rho_1, \xi} &= \exp \left( - \int_{0}^{\infty} \xd p \,  \frac{\pi }{4 \sinh (p \pi)} \left( \frac{I_{\im p}(m \rho_1)}{\overline{I_{\im |p|}(m \rho_1)} } 2 \xi(p) \xi(-p) + |\xi(p)|^2 + |\xi(-p)|^2 \right) \right).
\end{align}

The free amplitude for a coherent state results to be
\begin{align}
\varrho_{\rho_1}(K^S_{\rho_1,\xi}) = \int \xD \varphi \, \psi_{\rho_1, \xi} Z_{\rho_1,0}(\varphi) = \exp \left( -\frac{1}{2} \int_0^{\infty}  \xd p \, \frac{\pi}{2 \sinh(p \pi)} \left( |\xi(p)|^2 + |\xi(-p)|^2  + 2 \xi(p) \xi(-p) \right) \right),
\label{eq:freeamplM3}
\end{align}
which is independent of $\rho_1$, as it should.


\section{Identification of states}
\label{sec:identification}

In the last section we derived all the objects necessary for the GBF on $M_2$. We will now establish an isomorphism between the states on the boundary $\Sigma_{\rho_1}$ and $\partial M_1$ using the coherent states. Since the coherent states form a dense subset in the respective Hilbert spaces, it suffices if we can identify them.

\subsection{Holomorphic representation}
We have for the amplitude for a generic region $M$ and a coherent state $K^h_{\tau}$ the expression:
\begin{equation}
\label{eq:freeamplitude}
 \varrho_M(K^h_{\tau}) = \exp\left(\frac{1}{4}g_{\partial M}(\hat\tau,\hat\tau)\right),
\end{equation}
with $\hat\tau=\tau^R-i\tau^I$ and $\tau^R,\tau^I\in L_{\tilde M}$ such that $\tau=\tau^R+J_{\partial M}\tau^I$. The reader can easily verify that (\ref{eq:freeamplitude}) coincides with (\ref{eq:freeamphol}).

For region $M_1$ we obtain for solutions $\phi,\phi'$ in $L_{\tilde M_1}$ that
\begin{equation}
 g_{\partial M_1}(\phi,\phi') =2\int_{0}^{\infty} \xd p\,\left(\phi(p)\overline{\phi'(p)}+c.c.\right)\,.
\end{equation}

For the solution $\phi_{\Sigma_{\rho_1}}$ in $L_{\tilde M_2}\subset L_{\Sigma_{\rho_1}}$ we obtain by projecting the solution $\phi$ to a neighborhood of $\Sigma_{\rho_1}$ with the decomposition (\ref{eq:boundparam}) and using relation (\ref{eq:besselrel})
the identities
\begin{eqnarray}\label{eq:projectedcoeffs}
 \nonumber \phi_{\Sigma_{\rho_1}}(p)=\phi(p)\,\quad\textrm{and} \quad
 \phi_{\Sigma_{\rho_1}}(-p)=\overline{\phi(p)}.
\end{eqnarray}
For the metric we find then
\begin{eqnarray}
g_{\Sigma_{\rho_1}}\left(\phi_{\Sigma_{\rho_1}},\phi'_{\Sigma_{\rho_1}}\right)&=&\int_{-\infty}^{\infty} \xd p \, \left( \phi_{\Sigma_{\rho_1}}(p) \overline{\phi'_{\Sigma_{\rho_1}}(p)} + c.c. \right)\\
&=& 2\int_{0}^{\infty} \xd p\, \left(\phi(p)\overline{\phi'(p)}+c.c.\right).
\end{eqnarray}
Hence, identifying the expression in equation (\ref{eq:freeamplitude}) for the amplitude in $M_2$ and $M_1$ is equivalent to the identification
\begin{equation}
  \hat\tau_{\Sigma_{\rho_1}}=\hat\tau_{\partial M_1},
\end{equation}
for the two different regions. For region $M_2$, let us define $\tau_{\tilde M_2}^R(p)$ and $\tau_{\tilde M_2}^I(p)$ such that
\be
\tau^R_{\Sigma_{\rho_1}}=\int_{0}^{\infty} \xd p\, \left(\tau^R_{\tilde M_2} (p)\phi_p(x)+c.c.\right), \qquad
\tau^I_{\Sigma_{\rho_1}}=\int_{0}^{\infty} \xd p\, \left(\tau^I_{\tilde M_2}(p) \phi_p(x)+c.c. \right).
\ee
Then we obtain with 
the action of the complex structure corresponding to $\Sigma_{\rho_1}$ the identity
\begin{eqnarray}
\nonumber \tau_{\Sigma_{\rho_1}}&=&\tau^R_{\Sigma_{\rho_1}}+J_{\Sigma_{\rho_1}}\tau^I_{\Sigma_{\rho_1}}\\
&=&\int_{0}^{\infty} \xd p \,\left[\left(\tau^R_{\tilde M_2}(p)- \im \tau^I_{\tilde M_2}(p)\right)\chi_p(x)+\left(\tau^R_{\tilde M_2}(p)+ \im \tau^I_{\tilde M_2}(p)\right)\overline{\chi_{-p}(x)}+c.c.\right].
\end{eqnarray}
By comparing this with Equation (\ref{eq:boundparam}) (replacing $\phi_{\Sigma_{\rho_1}}$ by $\tau$) we obtain
\begin{eqnarray}
\tau_{\Sigma_{\rho_1}}(p) = \tau^R_{\tilde M_2}(p)- \im \tau^I_{\tilde M_2}(p), \qquad \overline{\tau_{\Sigma_{\rho_1}}(-p)} = \tau^R_{\tilde M_2}(p)+ \im \tau^I_{\tilde M_2}(p),
\end{eqnarray}
for $p>0$, which can be inverted as
\begin{eqnarray}
\tau^R_{\tilde M_2}(p)=\frac12\left(\tau_{\Sigma_{\rho_1}}(p)+\overline{\tau_{\Sigma_{\rho_1}}(-p)}\right), \qquad
 \tau^I_{\tilde M_2}(p)=\frac{\im}{2}\left(\tau_{\Sigma_{\rho_1}}(p)-\overline{\tau_{\Sigma_{\rho_1}}(-p)}\right).
\end{eqnarray}
Then we find the expression
\be
\hat\tau_{\Sigma_{\rho_1}} = \int_{0}^{\infty} \xd p\,\left[ \tau_{\Sigma_{\rho_1}}(p) \phi_p(x) + \tau_{\Sigma_{\rho_1}}(-p) \overline{\phi_p(x)}\right].
\ee
For region $M_1$ we have for a solution $(\tau_1,\tau_2)\in L_{\Sigma_1}\oplus L_{\overline{\Sigma_2}}=L_{\partial M_1}$ that $\tau^R=1/2(\tau_1+\tau_2,\tau_1+\tau_2)$ and $J_{\partial M_1}\tau^I=1/2(\tau_1-\tau_2,\tau_2-\tau_1)$ and hence $\hat\tau=1/2(1+ \im J_{\Sigma_1}) \tau_1+ 1/2(1- \im J_{\Sigma_1})\tau_2$ and we obtain
\begin{equation}
 \hat{\tau}_{\partial M_1}(x) = \int_0^{\infty} \xd p\, \left( \phi_p(x) \tau_{1}(p) +  \overline{\phi_p(x)} \overline{\tau_{2}(p)} \right)\,,
\end{equation}
which leads to the identification
\be
 \tau_1(p) = \tau_{\Sigma_{\rho_1}}(p), \qquad \overline{\tau_2(p)} = \tau_{\Sigma_{\rho_1}}(-p),
\ee
with $p>0$.
These expressions give an isomorphism between the Hilbert spaces on the boundary of $M_2$ and $M_1$. In particular, this isomorphism preserves the amplitude by construction and, thus, preserves the probability for the quantum field theory. It also preserves the vacuum state since $\psi_{0;\Sigma_{h}}=K_{0;\Sigma_{h}}$ is mapped to $\psi_{0;\partial M_{\eta}}=K_{0;\partial M_{\eta}}$.

We will show in Section \ref{sec:propeq} that also the observable amplitudes for certain Weyl observables of the form $W=\exp(\im D)$ with $D(\phi)=\int \xd^2 x \sqrt{-\det g(x)} \mu(x) \phi(x)$ and $\mu(x)$ a general test function are preserved. Since the corresponding amplitude can be used as a generating functional for the perturbative quantization of interacting scalar field theories, this means that the amplitudes for interacting scalar field theories in the two regions and are equivalent.

\subsection{Schr\"odinger-Feynman representation}
\label{sec:iso}

In Schr\"odinger-Feynman representation we proceed in a way analogous to what we did in the holomorphic representation. Based on previous results \cite{CoOe:letter,CoOe:Smatrix,CoOe:2d}, and in particular according to formula (75) of \cite{CoDo:gen}, in the region $M_1$ we have 
\begin{equation}
\hat{\xi}(\rho,\eta) = - \frac{\im }{2 p} \left( e^{- \im p \eta} \xi_1(\rho) + e^{\im p \eta} \overline{\xi_2(\rho)} \right), 
\label{eq:xihatM1}
\end{equation}
where $\frac{e^{\pm \im p \eta}}{2 p}$ is to be understood as an operator; expanding the function $\xi_{1,2}(\rho)$ according to (\ref{eq:expansionK}) 
we get
\begin{equation}
\hat{\xi}(\rho,\eta) = - \im \int_0^{\infty} \frac{\xd p}{2 p} \frac{\sqrt{2 p \sinh(\pi p)}}{\pi} K_{\im p}(m \rho) \left( e^{- \im p \eta} \xi_1(p) + e^{\im p \eta} \overline{\xi_2(p)} \right).
\label{eq:xihatM1-2}
\end{equation}
On the other hand, in the region $M_2$, according to formula (91) of \cite{CoDo:gen}, we have 
\begin{equation}
\hat{\xi}(\rho,\eta) = - K_{\im |p|}(m \rho) \xi(\eta), 
\end{equation}
where $K_{\im p}(m \rho)$ is to be understood as an operator; the substitution of $\xi(\eta)$ with its expansion $\xi(\eta)= \int \frac{\xd p}{\sqrt{2 \pi}} e^{\im p \eta} \xi(p)$ leads to
\begin{align}
\hat{\xi}(\rho,\eta) &= - \int_{-\infty}^{\infty} \frac{\xd p}{\sqrt{2 \pi}} K_{\im |p|}(m \rho) e^{\im p \eta} \xi(p) = - \int_{0}^{\infty} \frac{\xd p}{\sqrt{2 \pi}} K_{\im p}(m \rho) \left( e^{\im p \eta}\xi(p) + e^{-\im p \eta}\xi(-p) \right).
\label{eq:xihatM3}
\end{align}
Identifying (\ref{eq:xihatM1-2}) with (\ref{eq:xihatM3}) leads to the following relations, valid for $p>0$,
\begin{equation}
\xi(p) = \im \sqrt{\frac{\sinh(\pi p)}{\pi p}} \overline{\xi_2(p)}, \qquad \text{and} \qquad \xi(-p) = \im \sqrt{\frac{\sinh(\pi p)}{\pi p}} \xi_1(p).
\label{eq:iso}
\end{equation}
Then, the substitution of these expressions for $\xi(\pm p)$ in the free amplitude (\ref{eq:freeamplM3}) in region $M_2$ reduces to the free amplitude (\ref{eq:freeamplM1}) in region $M_1$. It must be noted that the isomorphism implemented by (\ref{eq:iso}) results to be an isometric isomorphism.

\subsubsection{Equivalence of states on the boundary of Rindler space}

Consider the vacuums state (\ref{eq:vacuum2-2}) defined on the hyperbola.
We notice that the surface of constant $\rho$ in the limit where $\rho$ tends to zero approaches the union of the surfaces defined by $\eta \rightarrow -\infty$ and $\eta \rightarrow +\infty$.\footnote{The Hilbert spaces associated to these hypersurfaces will be denoted as $\cH_{-\infty}$ and $\cH_{\infty}$ respectively.} It is then to be expected that the vacuum state (\ref{eq:vacuum2-2}) at $\rho=0$ reduces to the tensor product of two vacuum states (\ref{eq:vacuum1}) for $\eta \rightarrow -\infty$ and $\eta \rightarrow +\infty$, that implies that the operator appearing in the exponential of (\ref{eq:vacuum2-2}) tends to the one in the exponential of (\ref{eq:vacuum1}). This can be easily checked by the asymptotic property (\ref{eq:asympIK0}) of the modified Bessel function $I_{\im |p|}$,
\begin{equation}
\lim_{\rho \rightarrow 0} \im \rho\frac{d}{d\rho}\ln\left(\overline{I_{\im |p|}(m{\rho})}\right) =  |p|,
\end{equation}
which is indeed the operator characterizing the vacuum state (\ref{eq:vacuum1}). The normalization factor $C_{\rho_1}$ appearing in (\ref{eq:vacuum2-2}) satisfies
\be
|C_{\rho_1}|^2 
= \det \left( - \frac{\im }{2 \pi} \rho_1\frac{\xd}{\xd \rho_1}\ln\left( I_{\im |p|}(m{\rho_1})\right) + \frac{\im }{2 \pi} \rho_1 \frac{\xd}{\xd \rho_1} \ln\left(\overline{I_{\im |p|}(m{\rho_1})}\right)\right)^{1/2} 
=\det \left(  \frac{ 1}{\pi^2} \, \frac{ \sinh (|p| \pi)}{|I_{\im |p|}(m{\rho_1})|^2}\right)^{1/2}.
\ee
In the limit $\rho_1 \rightarrow 0$, using (\ref{eq:asympIK0}) we have that
\begin{equation}
|I_{\im |p|}(m \rho_1)|^2 \sim |\Gamma(\im |p| +1)|^{-2} = |\im |p| \Gamma(\im |p|)|^{-2} = \frac{ \sinh (\pi |p|)}{\pi |p|}.
\end{equation}
The modulus square of the normalization factor $C_{\rho_1}$, in the limit $\rho \rightarrow 0$ can then be written as
\begin{equation}
|C_{\rho_1}|^2 = \det \left( \frac{|p|}{\pi} \right)^{1/2},
\end{equation} 
and the vacuum state reads in this limit
\begin{align}
\psi_{{\rho_1 \rightarrow 0},0}(\varphi_0) = \det \left( \frac{|p|}{\pi} \right)^{1/4} e^{\im \, \text{arg}(C_{\rho_1})} \exp \left( - \int_0^{\infty} \xd p \, \varphi_0(p) p \varphi_0(-p) \right). 
\end{align}
In order for this state to correspond to the state $\psi_{\eta \rightarrow -\infty, 0} \otimes \overline{\psi_{\eta \rightarrow \infty, 0}} \in {\cH_{-\infty} \otimes \cH_{\infty}^*}$,
\begin{equation}
\psi_{\eta \rightarrow -\infty, 0}(\varphi_{-\infty}) \otimes \overline{\psi_{\eta \rightarrow \infty, 0}(\varphi_{\infty})} = \det \left( \frac{p}{\pi} \right)^{1/4} \exp \left( - \frac{1}{2} \int_0^{\infty} \xd p \left[ \varphi_{\infty}(p) p \varphi_{\infty}(p)  + \varphi_{-\infty}(p) p \varphi_{-\infty}(p) \right] \right),
\end{equation}
the following equality must be satisfied,
\begin{equation}
\frac{1}{2} \left[ \varphi_{\infty}(p) \varphi_{\infty}(p)  + \varphi_{-\infty}(p) \varphi_{-\infty}(p) \right] = \varphi_0(p) \varphi_0(-p).
\label{eq:varphilimit}
\end{equation}
With this equality, which relates the coefficient of the modes expansion of the field in the asymptotic hypersurfaces $\eta \rightarrow \pm \infty$ and $\rho_1 \rightarrow 0$, it can be shown that also asymptotic coherent states coincide, namely
\be
\psi_{\rho_1 \rightarrow 0,\xi}(\varphi_0) = \psi_{\eta \rightarrow -\infty, \xi_1} (\varphi_{-\infty}) \otimes \overline{\psi_{\eta \rightarrow \infty, \xi_2} (\varphi_{\infty})},
\ee
where $\psi_{\rho_1 \rightarrow 0,\xi} \in \cH_{\rho_1 \rightarrow 0}$ and $\psi_{\eta \rightarrow -\infty, \xi_1} \otimes \overline{\psi_{\eta \rightarrow \infty, \xi_2}} \in {\cH_{-\infty} \otimes \cH_{\infty}^*}$.


\subsubsection{Equivalence of probability}

In this section, we show how the probability computed in the two regions $M_1$ and $M_2$ are related. In the GBF, probabilities can be computed from the amplitude maps, and are encoded in the formula
\begin{equation}
P({\cal A}/ {\cal S}) = \frac{\langle \varrho_M \diamond P_{\cal S}, \varrho_M \diamond P_{\cal A}\rangle}{\langle \varrho_M \diamond P_{\cal S}, \varrho_M \diamond P_{\cal S}\rangle},
\label{eq:prob}
\end{equation} 
where $\cal A$ and $\cal S$ are subspaces of the Hilbert space $\cH_{\partial M}$ associated to the boundary $\partial M$ of the region $M$, and $P_{\cal A}$ and $P_{\cal S}$ the orthogonal projectors onto these subspaces. The symbol $\diamond$ denotes composition of maps. Consequently, $\varrho_M \diamond P_{\cal S}$ and $\varrho_M \diamond P_{\cal A}$ are linear maps from $\cH_{\partial M}$ to the complex numbers. Two conditions must be required for this composition: (i) the maps $\varrho_M \diamond P_{\cal S}$ and $\varrho_M \diamond P_{\cal A}$ are continuous and (ii) the map $\varrho_M \diamond P_{\cal S}$ does not vanish. Then, these maps can be viewed as elements in the dual Hilbert space $\cH^*_{\partial M}$ and the inner product $\langle \cdot, \cdot \rangle$ appearing in (\ref{eq:prob}) is the inner product of this dual Hilbert space.
$P({\cal A}/ {\cal S})$ represents the conditional probability for observing $\cal A$ given that $\cal S$ has been prepared.

We consider first the region $M_1$. In this case, there exists a natural decomposition of the boundary Hilbert space $\cH_{\partial M_1}$, namely $\cH_{\partial M_1} = \cH_1 \otimes \cH_2^*$. We can then choose the subspaces ${\cal S}_{M_1}$ and ${\cal A}_{M_1}$ as
\begin{equation}
{\cal S}_{M_1} = \{ \psi \otimes \xi : \xi \in \cH_2^* \} \qquad \text{and} \qquad {\cal A}_{M_1} = \{ \psi \otimes \xi : \psi \in \cH_1\}.
\end{equation}
In order to evaluate the numerator and denominator of (\ref{eq:prob}) it is convenient to introduce an ON-basis of the boundary Hilbert space $\cH_{\partial M_1}$. In particular, since $\cH_{\partial M_1}$ decomposes as the tensor product $\cH_1 \otimes \cH_2^*$, we introduce two ON-bases $\{\nu^1_k \}$ and $\{\nu^2_k \}$ for the spaces $\cH_1$ and $\cH_2$ respectively. Then, we have 
\begin{align}
\langle \varrho_M \diamond P_{{\cal S}_{M_1}}, \varrho_M \diamond P_{{\cal A}_{M_1}}\rangle &=  \sum_{k,l} \overline{\varrho_{M_1} \diamond P_{{\cal S}_{M_1}} (\nu^1_k \otimes \nu^2_l)}  \, \varrho_{M_1} \diamond P_{{\cal A}_{M_1}} (\nu^1_k \otimes \nu^2_l)  \nonumber\\
\langle \varrho_M \diamond P_{{\cal S}_{M_1}}, \varrho_M \diamond P_{{\cal S}_{M_1}}\rangle &=  \sum_{k,l} |\varrho_{M_1} \diamond P_{{\cal S}_{M_1}} (\nu^1_k \otimes \nu^2_l)|^2.
\end{align}
Without loss of generality, we can choose $\nu_1^1=\psi$ and $\nu_1^2 = \xi$, and the probability (\ref{eq:prob}) takes the form
\begin{equation}
P({\cal A}_{M_1}/ {\cal S}_{M_1}) = \frac{|\varrho_{M_1}(\psi \otimes \xi)|^2}{\sum_{l} |\varrho_{M_1}(\psi \otimes \nu^2_l)|^2}.
\label{eq:prob2}
\end{equation}
Also without loss of generality, we can choose the states $\psi$ and $\xi$ to be coherent states that we denote $K_{\xi_1}$ and $\overline{K_{\xi_2}}$, respectively:
\begin{equation}
P({\cal A}_{M_1}/ {\cal S}_{M_1}) = \frac{|\varrho_{M_1}(K_{\xi_1} \otimes \overline{K_{\xi_2}})|^2}{\sum_{l} |\varrho_{M_1}(K_{\xi_1} \otimes \nu^2_l)|^2}.
\label{eq:prob3}
\end{equation}
In order to give a more useful expression of the denominator, we use the resolution to identity provided by the coherent states to obtain:
\begin{equation}
\sum_{l} |\varrho_{M_1}(K_{\xi_1} \otimes \nu^2_l)|^2 = \sum_{l} \left| D^{-1} \int  \xd \zeta \, \xd \overline{\zeta}  \, \overline{C_{\nu_k^2,\zeta}} \,\, \varrho_{M_1}(K_{\xi_1} \otimes \overline{K_{\zeta}}) \right|^2,
\end{equation}
where $\overline{C_{\nu_k^2,\zeta}} = \langle \nu_k^2, K_{\zeta} \rangle_{\cH_2}$ and $D$ is the coefficient appearing in the resolution of the identity satisfied by the coherent states \cite{CoDo:gen}.
The isomorphism expressed by the relations (\ref{eq:iso}) can be used to map the subspaces ${\cal A}_{M_1}$ and  ${\cal S}_{M_1}$ of the Hilbert space associated to the boundary of the region $M_2$ to the corresponding subspaces ${\cal A}_{M_2}$ and  ${\cal S}_{M_2}$ defined for the theory in the region $M_2$. In particular, as we have seen, that the relations in equation (\ref{eq:iso}) transform the free amplitude $\varrho_{M_1}(K_{\xi_1} \otimes \overline{K_{\xi_2}})$ into the free amplitude $\varrho_{M_2}(K_{\xi})$; moreover the number $\overline{C_{\nu_k^2,\zeta}}$ is invariant under the action of the isometric isomorphism (\ref{eq:iso}). We can consequently conclude that the probabilities computed in the region $M_1$ for the free theory are the same as the one computed in the region $M_2$,
\begin{equation}
P({\cal A}_{M_1}/ {\cal S}_{M_1})\big|_{\hat{(\xi_1,\xi_2)}=\hat{\xi}} = P({\cal A}_{M_2}/ {\cal S}_{M_2}).
\end{equation}

{\bf I interchanged $\xi$ with $\zeta$ to fit equation \ref{eq:iso}.}


\section{Preservation of amplitudes in the interacting theory}
\label{sec:propeq}

In this section we will first compare the observable amplitude for Weyl observables $W(\phi)=\exp(iD(\phi))$ with $D(\phi)=\int \xd^2 x \sqrt{-\det g(x)} \mu(x)\phi(x)$, where $\mu(x)$ is a general test function in the regions $M_1$ and $M_2$. 

\subsection{Holomorphic representation}

For a general region $M$ we have from Proposition 4.3 of \cite{Oe:SFobs} the following expression for the observable amplitude:
\begin{eqnarray}\label{eq:muamp}
\varrho_M^W \left(K_\tau\right) &=& \varrho_M(K_\tau) \exp\left(\im \int_M \xd^2x\,\sqrt{-\det g(x)}\,\mu(x)\hat{\tau}(x)+\right.\nonumber\\
&&\left.+\frac{\im}{2}\int_M \xd^2 x \xd^2 x'\,\sqrt{\det g(x) \det g(x')}\,\mu(x)G^M_{F}(x,x')\mu(x')\right),
\end{eqnarray}
where $G^M_{F}(x,x')$ is the Feynman propagator constructed such that
\begin{equation}\label{eq:feynprobdef}
 (\eta_D-\im J_{\partial M}\eta_D)(x)=\int_M \xd^2x\,\sqrt{-\det g(x)}\,G^J_{F}(x,x')\mu(x')\,,
\end{equation}
where $\eta_D$ is the unique element of $J_{\partial M}L_{\tilde M}$ fulfilling the condition $D(\xi)=2\omega_{\partial M}(\xi,\eta_D)$ for all $\xi\in L_{\tilde M}$. 

Since we constructed the isomorphism between $\cH_{\Sigma_{\rho_1}}$ and $\cH_{\partial M_1}$ such that the expressions for $\hat{\tau}$ for the two regions coincide, we have that the observable maps coincide if the Feynman propagators coincide. In region $M_1$ we obtain for the Feynman propagator the following expression \cite{Colosi2012}:
\begin{align}\label{eq:feynrind}
 G^{M_1}_F(x,x') &= \im \int \xd p \left(\theta(\eta'-\eta)\overline{\phi_p^R(x)}\phi^R_p(x') + \theta(\eta-\eta') \phi^R_p(x) \overline{\phi^R_p(x')} \right), \nonumber\\
 &= \im \int_{0}^{\infty} \frac{\xd p}{2 p} \left( \theta(\eta'-\eta) e^{\im p(\eta - \eta')} + \theta(\eta-\eta') e^{\im p(\eta' - \eta)} \right) K_{\im p}(m \rho') K_{\im p}(m \rho) \frac{ p \sinh (p \pi)}{\pi^2}2.
\end{align}
For region $M_2$ we derive the Feynman propagator in the following: Let us assume that we are given a function $\phi_{\tilde M_2}\in L_{\tilde M_2}\subset L_{\Sigma_{\rho_1}}$ such that $J_{\Sigma_{\rho_1}}\phi_{\tilde M_2}=\eta_D$. Let us decompose $\phi_{\tilde M_2}$ as in equation (\ref{eq:intparam}). Then we find that
\begin{eqnarray}\label{eq:etad}
 \eta_D(x)&=&\int_{-\infty}^\infty \xd p \, \left(\eta_D(p)\chi_p(x) + c.c.\right),\nonumber\\
&=& -i \int_{0}^\infty \xd p \, \left(\phi_{\tilde M_2}(p)\left(\chi_p(x)-\overline{\chi_{-p}(x)}\right)-\overline{\phi_{\tilde M_2}(p)}\left(\overline{\chi_p(x)}-\chi_{-p}(x)\right)\right),
\end{eqnarray}
from which we obtain that $\eta_D(p)=- \im \phi_{\tilde M_2}(p)$ for $p>0$ and $\eta_D(p)=- \im \overline{\phi_{\tilde M_2}(-p)}$ for $p<0$. Hence, we have for $\xi\in L_{\tilde M_2}$ using the identities in equation (\ref{eq:projectedcoeffs}) that
\begin{eqnarray}
 \omega_{\Sigma_{\rho_1}}(\xi,\eta_D)&=&-\frac{\im}{2}\int_{0}^\infty \xd p \,\left(\xi_{\Sigma_{\rho_1}}(p) \im \overline{\phi_{\tilde M_2}(p)}+\xi_{\Sigma_{\rho_1}}(-p) \im \phi_{\tilde M_2}(p)-c.c.\right),\nonumber\\
 &=&\int_{0}^\infty \xd p \,\left(\xi(p)\overline{\phi_{\tilde M_2}(p)}+c.c.\right)
\end{eqnarray}
From the condition $D(\xi)=\int \xd \eta \xd \rho\rho \,\mu(x)\xi(x)=\omega_{\Sigma_{\rho_1}}(\xi,\eta_D)$ we obtain that
\begin{equation}
 \phi_{\tilde M_2}(p)=\int \xd \eta' \xd \rho' \rho' \mu(x') \overline{\phi_p(x')},
\end{equation}
and with equation (\ref{eq:etad}) we find an expression for $\eta_D$. Now we are interested in the projection of the Feynman propagator to the boundary $\eta_D- \im J_{\Sigma_{\rho_1}}\eta_D$. We obtain
\be
\eta_D - \im J_{\Sigma_{\rho_1}} \eta_D = (\im +J_{\Sigma_{\rho_1}})\phi_{\tilde M_2} = 2 \im \int_{0}^\infty \xd p \, \left(\phi_{\tilde M_2}(p) \, \overline{\chi_{-p}(x)} + \overline{\phi_{\tilde M_2}(p)} \, \overline{\chi_p(x)} \right).
\ee
Using that $\phi_p(x)=\overline{\phi_{-p}(x)}$, we find for the Feynman propagator the symmetrized expression
\begin{align}
 G^{M_2}_F(x,x') &= \im \int_{-\infty}^\infty \xd p \, \left[ \theta(\rho'-\rho) \overline{\chi_{-p}(\eta,\rho)} \, \overline{\phi_p(\eta',\rho')}  + \theta(\rho-\rho')  \overline{\chi_{-p}(\eta',\rho')} \, \overline{\phi_p(\eta,\rho)}\right], \nonumber\\
 &=  \int_{-\infty}^{\infty} \frac{\xd p}{2 \pi} \left[ \theta(\rho'-\rho) K_{\im |p|}(m \rho') \overline{I_{\im |p|}(m \rho)} + \theta(\rho-\rho') K_{\im |p|}(m \rho) \overline{I_{\im |p|}(m \rho')} \right] e^{\im p (\eta-\eta')}.
\label{eq:FeprM3h}
\end{align}

\subsection{Schr\"odinger-Feynman quantization}
\label{sec:int}
A way to compute the expectation value of the Weyl observable $W$ is to modify the action as
\begin{equation}
S_{M,\mu}(\phi) = S_{M,0}(\phi) + \int_M \xd^2 x \sqrt{- \det g(x)} \, \phi(x) \, \mu(x).
\label{eq:acint}
\end{equation}
The form of the corresponding field propagator (\ref{eq:proppint0}) can be obtained by 
shifting the integration variable by a classical solution $\phi_{\text{cl}}$ that matches the boundary configuration $\varphi$ on the boundary $\partial M$,
\be
Z_{M,\mu}(\varphi) = \int_{\phi|_{\partial M} =\varphi} \xD \phi \, e^{\im S_{M,\mu}(\phi)} = \int_{\phi|_{\partial M} =0} \xD \phi \, e^{\im S_{M,\mu}(\phi_{\text{cl}} + \phi)} = N_{M,\mu} \, e^{\im S_{M,\mu}(\phi_{\text{cl}})},
\ee
where $N_{M,\mu} = \int_{\phi|_{\partial M} =0} \xD \phi \, e^{\im S_{M,\mu}(\phi)}.$
The propagator can be expressed in terms of the propagator $Z_{M,0}(\varphi)$ of the free theory as
\be
Z_{M,\nu}(\varphi) = Z_{M,0}(\varphi) \, \exp \left( \im \int_M \xd^2 x \sqrt{- \det g(x)} \, \phi_{\text{cl}} \, \mu(x) +\frac{\im}{2} \int_M \xd^2 x \sqrt{- \det g(x)} \, \alpha(x) \, \mu(x) \right),
\label{eq:propnu}
\ee
where the quantity $\alpha$ is the solution of the inhomogeneous equation $\left(- \rho \partial_\rho \rho \partial_\rho + \partial^2_\eta + m^2 \rho^2 \right) \alpha(\eta, \rho) = \mu(\eta,\rho)$,
with vanishing boundary condition $\alpha|_{\partial M}=0$.
In the region $M_1$, a classical solution with boundary configurations $\varphi_1$ and $\varphi_2$ is given by (\ref{eq:phiM1}) and the function $\alpha$ results to be
\be
\alpha(\eta,\rho) =\int_{\eta_1}^{\eta_2} \xd \eta' \, \rho \left( \theta(\eta'-\eta) \frac{\sin p(\eta-\eta_1) \sin p(\eta_2-\eta')}{p \sin p(\eta_2-\eta_1)} + \theta(\eta-\eta') \frac{\sin p(\eta'-\eta_1) \sin p(\eta_2-\eta)}{p \sin p(\eta_2-\eta_1)} \right).
\ee
Notice that $\alpha(\eta_1,\rho)=\alpha(\eta_2,\rho)=0$.
Substituting these quantities in the expression of the propagator (\ref{eq:propnu}) and performing the integration in (\ref{eq:amplitude}) leads to the amplitude for a coherent state $ K^S_{\eta_1,\xi_1} \otimes \overline{K^S_{\eta_2,\xi_2}}$
\begin{multline}
\varrho_{[\eta_1,\eta_2]}^{W} \left( K^S_{\eta_1,\xi_1} \otimes \overline{K^S_{\eta_2,\xi_2}} \right) = \varrho_{[\eta_1,\eta_2]} \left( K^S_{\eta_1,\xi_1} \otimes \overline{K^S_{\eta_2,\xi_2}} \right) \exp \left( \int_{M_1} \xd^2 x \, \sqrt{-g(x)} \, \hat{\xi}(x) \mu(x) \right)\\ 
\times \exp \left( \frac{\im}{2} \int_{M_1} \xd^2 x \, \xd^2 x' \, \sqrt{g(x)g(x')} \mu(x) G_F^{M_1}(x,x') \mu(x')\right),
\label{eq:amplM1mu-1}
\end{multline}
where $\varrho_{[\eta_1,\eta_2]} \left( K^S_{\eta_1,\xi_1} \otimes \overline{K^S_{\eta_2,\xi_2}} \right)$ is the free amplitude (\ref{eq:freeamplM1}), $\hat{\xi}$ is the complex solution given by (\ref{eq:xihatM1}) and $G_F^{M_1}(x,x')$ is the Feynman propagator in region $M_1$ given by (\ref{eq:feynrind}). Taking the limit $\eta_1 \rightarrow -\infty$ and $\eta_2 \rightarrow +\infty$ in the amplitude (\ref{eq:amplM1mu-1}) reduces to substitute the subindex $M_1$ with the whole Rindler space.

In the region $M_2$, a classical solution with boundary configuration $\varphi$ is given by (\ref{eq:phiM3}) and
$\alpha$ can be expressed in integral form as $\alpha(\eta,\rho) = \int_{\rho_1}^{\infty} \xd \rho' \, \rho' \, \tilde{g}(\rho,\rho') \,  \nu(\eta, \rho')$, where
\begin{equation}\label{eq:gfunction}
\tilde{g}(\rho,\rho') = - \theta(\rho'-\rho) \left( L_{\im p}(m\rho') K_{\im p}(m\rho) - L_{\im p}(m\rho) K_{\im p}(m\rho') \right) + L_{\im p}(m\rho') K_{\im p}(m\rho) - K_{\im p}(m\rho) \frac{L_{\im p}(m\rho_1)}{K_{\im p}(m\rho_1)} K_{\im p}(m\rho'),
\end{equation}
where $L_{\im p}$ is the real part of $I_{\im p}$.
Notice that $\alpha$ satisfied the vanishing boundary condition $\alpha(\eta,\rho_1)=0$.
The expression for the amplitude of a coherent state in the interacting theory results to be
\begin{multline}
\varrho_{M_2}^{W}(K^S_{\rho_1, \xi}) = \varrho_{M_2}(K^S_{\rho_1, \xi}) \exp \left( \int_{M_2} \xd^2 x \, \sqrt{-g(x)} \, \hat{\xi}(x) \mu(x) + \frac{\im}{2} \int_{M_2} \xd^2 x \, \xd^2 x' \, \sqrt{g(x)g(x')} \mu(x) G_F^{M_2}(x,x') \mu(x')\right),
\label{eq:amplM3nu-1}
\end{multline}
where $x$ is a global notation for the coordinates $\eta,\rho$ and $G_F^{M_2}(x,x')$ is given by (\ref{eq:FeprM3h}).
Taking the limit $\rho_1 \rightarrow 0$ in the amplitude (\ref{eq:amplM3nu-1}) reduces to substitute the subindex $M_2$ with the whole Rindler space.


\subsection{Equality of the Feynman propagators in region $M_1$ and $M_2$}
\label{app:Feyn}

In this section, we show in two different ways the equality of the propagators in the region $M_1$ and $M_2$, i.e. we show the identity $G^{M_2}_F(x,x')=G^{M_1}_F(x,x')$.
This result means that the observable amplitudes $\varrho_{M_1}^W(\Psi)$ and $\varrho_{M_2}^W(\Psi')$ coincide for all Weyl observables of the form $W(\phi)=e^{iD(\phi)}$ with $D(\phi)=\int \xd^2 x \sqrt{-\det g(x)} \, \mu(x) \phi(x)$ when the state $\Psi$ is mapped to $\Psi'$ with the isomorphism we identified in Section \ref{sec:identification} and $\mu(x)$ has support in the interior of both regions. These amplitudes can be used as generating functionals to derive all the n-point functions of the field $\phi$ which, thus, also coincide for the two regions. For a quantum field theory of two interacting scalar fields $\phi_1$ and $\phi_2$, the corresponding amplitude can also be generated using the amplitude in equation (\ref{eq:muamp}) as a generating functional \cite{CoOe:letter}. Hence, the coincidence of the vacuum state, amplitudes and probabilities is also valid for the interacting theory.

\subsubsection{First method}

We start from expression (\ref{eq:feynrind}) of the Feynman propagator in region $M_1$. 
The integral can be extended to negative values of $p$ by substituting $p$ with $|p|$; then, using the relation
\begin{equation}
\frac{\im}{2 |p|} \left( \theta(\eta'-\eta) e^{\im |p|(\eta - \eta')} + \theta(\eta-\eta') e^{\im |p|(\eta' - \eta)} \right) =
-  \lim_{\epsilon \rightarrow 0} \int_{-\infty}^{\infty} \frac{\xd q}{2\pi} \frac{e^{- \im q (\eta-\eta')}}{q^2-p^2+ \im \epsilon},
\end{equation}
and expressing the Macdonald function in terms of the modified Bessel functions of the first kind, (\ref{eq:besselrel}),
we obtain
\begin{align}\label{eq:propM1-2}
G_F^{M_1}(x,x') &= \frac{1}{4}  \int_{-\infty}^{\infty} \frac{\xd q}{2 \pi} \int_{-\infty}^{\infty} \xd p \, 
\frac{e^{- \im q (\eta-\eta')}}{q^2-p^2+ \im \epsilon} (I_{- \im p}(m \rho') - I_{\im p}(m \rho'))  (I_{- \im p}(m \rho) - I_{\im p}(m \rho))
 \frac{p}{ \sinh (p \pi)}, \nonumber\\
&= \int_{-\infty}^{\infty} \frac{\xd q}{2 \pi} e^{- \im q (\eta-\eta')} \left( \mathcal{I}_{++} + \mathcal{I}_{--} -\mathcal{I}_{-+} -\mathcal{I}_{+-} \right),
\end{align}
where we introduced the notation
\begin{equation}
\mathcal{I}_{lm} =  \frac{1}{4} \int_{-\infty}^{\infty} \xd p \, \frac{1}{q^2-p^2+ \im \epsilon} \, \frac{p}{ \sinh (p \pi)} \, I_{l \im p}(m \rho') I_{m \im p}(m \rho), \qquad (l=+,-), (m=+,-).
\end{equation}
In the following, we will perform the integration over $p$ for every term  $\mathcal{I}_{lm}$ with $l,m=\pm 1$ separately. First of all, we notice that the each term $\mathcal{I}_{lm}$ apparently contains an infinite number of poles for $p= \im n$, where $n$ is an integer. However, it can be shown that only the two poles $p_{\pm}= \pm (|q|+ \im \epsilon)$ contribute to the sum in (\ref{eq:propM1-2}). We apply the complex contour integration to evaluate their contribution. We start with the integral $\mathcal{I}_{++}$ which is equal to
\begin{equation}
\mathcal{I}_{++} = - \frac{1}{4} \int_{-\infty}^{\infty} \xd p \,  \frac{1}{p^2-q^2- \im \epsilon} \frac{p}{ \sinh{p \pi}}  I_{\im p}(m \rho) I_{\im p}(m \rho').
\end{equation}
We rewrite this integral using formula (5.7.1) of \cite{lebedev1972},
\begin{equation}
I_{\nu}(z) = \sum_{k=0}^{\infty} \frac{(z/2)^{\nu+2k}}{\Gamma(k+1) \Gamma(k+\nu+1)},
\end{equation}
which is valid for $|z| < \infty$, $|\text{arg} \, z|<\pi$. Substituting the above expression in $\mathcal{I}_{++}$, we get
\begin{equation}
\mathcal{I}_{++} = - \frac{1}{4} \int_{-\infty}^{\infty} \xd p \,  \frac{1}{p^2-q^2- \im \epsilon} \frac{p}{ \sinh{p \pi}}  
\sum_{k,k'=0}^{\infty} \frac{(m \rho/2)^{2k} (m \rho'/2)^{2k'}}{\Gamma(k+1) \Gamma(k'+1)} \, \frac{(m \rho/2)^{\im p} (m \rho'/2)^{\im p}}{\Gamma(k+1+\im p) \Gamma(k'+1+\im p)}.
\end{equation}
We compute this integral by closing the contour of integration in the complex $p$ plane. To do this we look at the behavior of the gamma functions for large values of the argument. We use the asymptotic expansion (1.4.23) of  \cite{lebedev1972},
\begin{equation}
\Gamma(z)= e^{(z-1/2) \log z - z +1/2 \log 2\pi} \left( 1+ O(|z|^{-1})\right),
\end{equation}
which is valid for $|\text{arg}\,z|<\pi$. Substituting in $\mathcal{I}_{++}$ we get
\begin{align}
\mathcal{I}_{++} \approx& - \frac{1}{4} \int_{-\infty}^{\infty} \xd p \,  \frac{1}{p^2-q^2- \im \epsilon} \frac{p}{ \sinh{p \pi}}  
\sum_{k,k'=0}^{\infty} \frac{(m \rho/2)^{2k} (m \rho'/2)^{2k'}}{\Gamma(k+1) \Gamma(k'+1)}\times \nonumber\\
& \times\exp \left( \im p (\log(m^2 \rho \rho'/4) - \log(k+1+\im p) - \log(k'+1+\im p) +2 )\right)\times \nonumber\\
& \times\exp \left( - (k+1/2)\log(k+1+\im p) - (k'+1/2)\log(k'+1+\im p) - \log(2 \pi) +k+k'+2 \right)
\label{eq:I++}
\end{align}
We write $p= r e^{\im \theta}$, consequently
\begin{align}
\log(k+1+\im p) &= \log(k+1+\im r e^{\im \theta}) = \log(k+1+\im r \cos \theta - r \sin \theta) \nonumber\\
&= \log \sqrt{(k+1-r \sin\theta)^2 + r^2 \cos^2 \theta} + \im \arctan \frac{r \cos \theta}{k+1-r \sin \theta} \nonumber\\
&= \log \sqrt{(k+1)^2- 2(k+1)r \sin\theta + r^2 } + \im \arctan \frac{r \cos \theta}{k+1-r \sin \theta}
\end{align}
which for $r >> (k+1)$ reduces to $\log(k+1+\im p) \approx \log r + \im \arctan \left( - \cot \theta \right)$.
Then we have that the argument of the first exponential in (\ref{eq:I++}) can be rewritten as
\begin{align}
&\im p (\log(m^2 \rho \rho'/4) - \log(k+1+\im p) - \log(k'+1+\im p) +2 ) \nonumber\\
&= \im r e^{\im \theta} \left( \underbrace{\log(m^2 \rho \rho'/4) - 2 \log r + 2}_{\tilde{r}} -2 \im \arctan \left( - \cot \theta \right) 	\right) \nonumber\\
&= \im \left( r \tilde{r} \cos \theta  + 2 r \sin \theta \, \arctan \left( - \cot \theta \right) \right) - r \left(\tilde{r} \sin \theta - 2 \cos \theta \, \arctan \left( - \cot \theta \right)\right).
\end{align}
Let us have a close look at the factor in the last term:
\begin{equation}
 \tilde{r} \sin \theta - 2 \cos \theta \, \arctan \left( - \cot \theta \right)=\left(\log(m^2 \rho \rho'/4) - 2 \log r + 2\right)\sin \theta - 2 \cos \theta \, \arctan \left( - \cot \theta \right).
\end{equation}
For finite $\rho,\rho'$ and $\theta\in[-\pi,0]$ we can always choose $r$ large enough to get this factor positive. We find that we can close the contour of integration in the lower half plane, namely $\theta\in[-\pi,0]$, send $r\to \infty$ and apply the residue theorem. The pole in the lower half plane is located in $-|q|-\im \epsilon$ and the result of the integration is
\begin{equation}
\mathcal{I}_{++} = - \frac{1}{4} \im \frac{\pi}{ \sinh |q| \pi} I_{-\im |q|}(m\rho) I_{ -\im |q|}(m \rho').
\end{equation}
We obtain the same expression for $\mathcal{I}_{--}$, namely $\mathcal{I}_{++} =\mathcal{I}_{--}$.
For the integral $\mathcal{I}_{+-}$  and $\mathcal{I}_{-+}$, applying similar techniques we obtain
\begin{equation}
\mathcal{I}_{+-} = \mathcal{I}_{-+} = - \frac{1}{4} \im \, \frac{\pi}{ \sinh (|q| \pi)} \left( \theta( \rho - \rho' )  I_{\im |q|}(m \rho) I_{-\im |q|}(m \rho') + \theta( \rho' - \rho ) I_{-\im |q|}(m \rho) I_{\im |q|}(m \rho') \right).
\end{equation}
Finally, 
the Feynman propagator in the region $M_1$ results to be
\begin{align}
G_F^{M_1}(x,x')
&= \int_{-\infty}^{\infty} \frac{\xd q}{2 \pi} e^{- \im q (\eta-\eta')} \left[ \theta( \rho - \rho' )  K_{\im |q|}(m \rho) I_{-\im |q|}(m \rho') + \theta( \rho' - \rho ) I_{-\im |q|}(m \rho) K_{\im |q|}(m \rho') \right],
\end{align}
where relation (\ref{eq:besselrel}) has been used. This propagator coincides with the propagator (\ref{eq:FeprM3h}) 
in the region $M_2$, namely $\displaystyle{G_F^{M_1}(x,x') = G_F^{M_2}(x,x').}$


\subsubsection{Second method}
\label{sec:reexpress}

We consider formula 7.213 of \cite{GrRi},
\begin{equation}
\int_0^{\infty} \frac{x \tanh(\pi x)}{\alpha^2 + x^2} P_{-\frac{1}{2} + \im x} ( \cosh \beta) \, \xd x = Q_{\alpha - \frac{1}{2}} ( \cosh \beta), \qquad \Re(a)>0,
\end{equation}
where $P_n$ and $Q_n$ are the associated Legendre functions of the first and second kind respectively. We set $\alpha= \im \sqrt{p^2- \im \epsilon} \simeq \im |p| + \epsilon$, with $\epsilon >0$ and $\epsilon << 1$. So, 
\begin{equation}
\int_0^{\infty} \frac{x \tanh(\pi x)}{-|p|^2 + \im \epsilon' + x^2} P_{-\frac{1}{2} + \im x} \left( \frac{a^2+b^2+c^2}{2ab}\right) \, \xd x \simeq Q_{\im |p| + \epsilon - \frac{1}{2}} \left( \frac{a^2+b^2+c^2}{2ab}\right),
\label{eq:intLegendre}
\end{equation}
where we also have replaced $\cosh \beta$ with $\frac{a^2+b^2+c^2}{2ab}$, $\epsilon'=|p|\epsilon$ is still very small and equality holds for $\epsilon'\to0$. Consequently the above equation is valid for $\frac{a^2+b^2+c^2}{2ab} \geq 1$.
We now consider the formula 6.672.3 of \cite{GrRi},
\begin{equation}
\int_0^{\infty} K_{\nu}(ax) K_{\nu}(bx) \cos(cx) \xd x = \frac{\pi^2}{4 \sqrt{ab}} \sec(\pi \nu) \text{P}_{\nu - \frac{1}{2}} \left( \frac{a^2+b^2+c^2}{2ab} \right), \qquad \Re(a+b)>0, c>0, |\Re(\nu)|< \frac{1}{2}.
\label{eq:KK-P}
\end{equation}
We multiply by $\cos(c y)$, ($y>0$), both sides and then integrate with respect to $c$.
It is easy to show that the integrals in the l.h.s of (\ref{eq:KK-P}) result to be equal to $\frac{\pi}{2} K_{\nu}(ay) K_{\nu}(by)$.
Then
\begin{equation}
K_{\nu}(ay) K_{\nu}(by) = \frac{\pi}{2 \sqrt{ab}} \sec(\pi \nu) \int_0^{\infty} \xd c \, \cos(c y) \text{P}_{\nu - \frac{1}{2}} \left( \frac{a^2+b^2+c^2}{2ab} \right),
\label{eq:KK}
\end{equation}
which is valid for $y>0, \Re(a+b)>0, |\Re(\nu)|< \frac{1}{2}$.

We now consider the formula 6.672.4 of \cite{GrRi},
\begin{equation}
\int_0^{\infty} K_{\nu}(ax) I_{\nu}(bx) \cos(cx) \xd x = \frac{1}{2 \sqrt{ab}} \text{Q}_{\nu - \frac{1}{2}} \left( \frac{a^2+b^2+c^2}{2ab} \right), \qquad \Re(a)>|\Re(b)|, c>0, \Re(\nu)> - \frac{1}{2}.
\end{equation}
By applying the same technique, namely by multiplying by $\cos(c y)$, ($y>0$), both sides and then integrate with respect to $c$, we obtain
\begin{equation}
K_{\nu}(ay) I_{\nu}(by) = \frac{1}{\pi \sqrt{ab}} \int_0^{\infty} \xd c \, \cos(c y) \text{Q}_{\nu - \frac{1}{2}} \left( \frac{a^2+b^2+c^2}{2ab} \right),
\label{eq:IK}
\end{equation}
which is valid for $y>0, \Re(a)>|\Re(b)|, \Re(\nu)> - \frac{1}{2}$.

We multiply by $\cos(c y)$, ($y>0$), both sides of equation (\ref{eq:intLegendre}) and then integrate with respect to $c$,
\begin{equation}
\int_0^{\infty} \xd c \, \cos(c y) \int_0^{\infty} \frac{x \tanh(\pi x)}{-|p|^2 + \im \epsilon + x^2} P_{-\frac{1}{2} + \im x} \left( \frac{a^2+b^2+c^2}{2ab}\right) \, \xd x \simeq \int_0^{\infty} \xd c \, \cos(c y) Q_{\im |p| + \epsilon - \frac{1}{2}} \left( \frac{a^2+b^2+c^2}{2ab}\right),
\label{eq:intLegendre2}
\end{equation}
and invert the integral on the l.h.s. which leads to, using (\ref{eq:KK}) and (\ref{eq:IK})
\begin{equation}
\int_0^{\infty} \frac{x \tanh(\pi x)}{-|p|^2 + \im \epsilon + x^2} K_{\im x}(ay) K_{\im x}(by)  \frac{2 \sqrt{ab}}{\pi} \cos(\pi \im x) \, \xd x = 
\pi \sqrt{ab} K_{\im |p| + \epsilon}(ay) I_{\im |p| + \epsilon}(by),
\end{equation}
or equivalently
\begin{equation}
\int_0^{\infty} \frac{x \sinh(\pi x)}{-|p|^2 + \im \epsilon + x^2} K_{\im x}(ay) K_{\im x}(by)  \, \xd x = 
\frac{\pi^2}{2} \, K_{\im |p| + \epsilon}(ay) I_{\im |p| + \epsilon}(by),
\label{eq:relI-K}
\end{equation}
which is valid for $y>0, \Re(a+b)>0, \Re(a)>|\Re(b)|, \epsilon>0, \epsilon << 1$.

We now rewrite the Feynman propagator in the region $M_2$ (\ref{eq:FeprM3h}), as 
\begin{equation}
G_F^{M_2}(x,x') =  \lim_{\epsilon \rightarrow 0}\int_{-\infty}^{\infty} \frac{\xd p}{2 \pi} \left[ \theta(\rho'-\rho) K_{\im |p|+ \epsilon}(m \rho') \overline{I_{\im |p|+ \epsilon}(m \rho)} + \theta(\rho-\rho') K_{\im |p|+ \epsilon}(m \rho) \overline{I_{\im |p|+ \epsilon}(m \rho')} \right] e^{\im p (\eta-\eta')}.
\label{eq:FeprM3-2}
\end{equation}
We use the relation (\ref{eq:relI-K}) with the following identifications (which satisfy the conditions for the validity of (\ref{eq:relI-K}))
\begin{align}
y&=m>0,\\
a&=\rho, b=\rho', \qquad \text{for} \qquad \rho > \rho',\\ 
a&=\rho', b=\rho, \qquad \text{for} \qquad \rho' > \rho,
\end{align}
and obtain
\begin{equation}
G_F^{M_2}(x,x') =  \lim_{\epsilon \rightarrow 0} \int_{-\infty}^{\infty} \frac{\xd p}{2 \pi} \, \frac{2}{\pi^2} e^{\im p (\eta-\eta')}
\int_0^{\infty} \frac{x \sinh(\pi x)}{-|p|^2 - \im \epsilon + x^2} K_{\im x}(m \rho) K_{\im x}(m \rho')  \, \xd x.
\label{eq:FeprM3-3}
\end{equation}
We invert the order of integration and perform first the integral over $\xd p$. For $\eta > \eta'$ we close the contour of integration in the upper half plane and for $\eta < \eta'$ in the lower half plane; the poles are $p_{\pm}=\pm (|x| - \im \epsilon)$. We obtain
\begin{equation}
\int_{-\infty}^{\infty} \frac{\xd p}{2 \pi} \,  \frac{e^{\im p (\eta-\eta')}}{-p^2 - \im \epsilon + x^2} =   \frac{\im}{2(|x| - \im \epsilon)} \left[ \theta(\eta-\eta') e^{-\im (|x| - \im \epsilon) (\eta-\eta')} + \theta(\eta'-\eta) e^{\im (|x| - \im \epsilon) (\eta-\eta')} \right].
\end{equation}
The Feynman propagator takes the form
\begin{align}
G_F^{M_2}(x,x') &=   \lim_{\epsilon \rightarrow 0} \int_0^{\infty}  
\frac{\im}{2(|x| - \im \epsilon)} \left[ \theta(\eta-\eta') e^{- \im (|x| - \im \epsilon) (\eta-\eta')} + \theta(\eta'-\eta) e^{\im (|x| - \im \epsilon) (\eta-\eta')} \right]
K_{\im x}(m \rho) K_{\im x}(m \rho')  \, \frac{2x \sinh(\pi x)}{\pi^2} \xd x, \nonumber\\
&=  \int_0^{\infty}  
\frac{\im}{2x} \left[ \theta(\eta-\eta') e^{- \im x (\eta-\eta')} + \theta(\eta'-\eta) e^{\im x (\eta-\eta')} \right]
K_{\im x}(m \rho) K_{\im x}(m \rho')  \, \frac{2x \sinh(\pi x)}{\pi^2} \xd x,
\label{eq:FeprM3-4}
\end{align}
which coincides with the expression (\ref{eq:feynrind}) of the Feynman propagator in region $M_1$, $\displaystyle{G_F^{M_1}(x,x') =  G_F^{M_2}(x,x').}$


\section{Summary and outlook}
\label{sec:conclusions}

We constructed the general boundary quantum field theory for a scalar field in 2-dimensional Rindler space in two different regions: a region $M_1$ with spacelike boundaries  and a region $M_2$ with purely timelike boundary. More specifically, the boundary of region $M_1$ was given by the disjoint union of two equal Rindler time hypersurfaces and the boundary of region $M_2$ was given as a timelike curve of constant Rindler spatial coordinate. 
We showed the existence of an isomorphism between the Hilbert spaces associated with these boundaries.

The isomorphism we identified preserves the amplitude map, and thus, the probabilities that can be extracted from the free quantum field theories are also preserved. We showed that the amplitude is also preserved when an interaction of the quantum field with a classical source is considered. That was done by showing that the isomorphism preserves the generating functional for perturbative quantum field theory. To obtain this result we showed that the Feynman propagators for the quantum field theories in the two regions are equivalent. Consequently we have obtained two equivalent representations of the Feynman propagator in Rindler space. This generalizes previous results obtained for QFT in Rindler space \cite{Birrell:1982ix}.

In particular, the generating functional for a given source term is equivalent with the expectation value (operator amplitude \cite{Oe:Obs}) of a particular local Weyl observable associated with that source term. We concluded that the expectation values for these observables are also preserved under the action of the isomorphism we identified. 

Let us emphasize again that regions with timelike boundaries like $M_2$ cannot be considered in the standard formulation of quantum field theory. The case investigated in this article shows that pairs of regions exist in Rindler space where one of these regions has timelike boundaries and the other region has spacelike boundaries such that both regions can be used equivalently to describe the same physical situation. Analogous results have been obtained within the GBF in Minkowski space \cite{CoOe:letter,CoOe:Smatrix}, a Euclidean space \cite{CoOe:2d} and de Sitter space \cite{Co:letter,Co:dS}. In Minkowski space, this result was used to show explicitly that the crossing symmetry is generic in the GBF.

The result presented here will find an immediate application in the context of the so called Unruh effect which is often derived from a comparison between the QFT in Minkowski and Rindler spaces. From such a perspective, it is of particular interest that the region $M_2$ does not extend to the spacelike infinity of Rindler space at $\rho= 0$. If Rindler space is embedded in Minkowski space as the right Rindler wedge, this point is mapped to the origin of Minkowski space. The mathematical problems arising from the singular behavior of the mode expansions used for the derivation of the Unruh effect at the origin of Minkowski space led to a critique of the mathematical basis of the Unruh effect by Narozhnyi et.al. in \cite{Bel1997,Fedotov:1999gp,Narozhnyi:2000rh,Bel2001,Bel:QF-BH}\footnote{See also the answer by Fulling and Unruh in \cite{Fulling2004} and a reply by Narozhnyi et.al. in \cite{Narozhny2004}.}. By investigating the Unruh effect using region $M_2$ such problems would be completely avoided. Moreover, the hypercylinder region and isomorphism constructed between the Hilbert spaces used in the different regions can provide a new representation of the mixed state involved in the Unruh effect. This will offer the possibility to study the properties of such state from a novel perspective. We shall elaborate on that elsewhere.

\begin{acknowledgments}

The authors are grateful to Robert Oeckl for useful comments on an earlier draft of this paper. The work of DR has been supported by the International Max Planck Research School for Geometric Analysis, Gravitation and String Theory.

\end{acknowledgments}

\appendix
\section{Modified Bessel functions}
\label{sec:appA}

The modified Bessel function of the first kind $I_{\im p}$, with imaginary order, and the modified Bessel function of the second kind $K_{\im p}$, also known as Macdonald function, are related by \cite{GrRi}
\begin{equation}\label{eq:besselrel}
 K_{ip}=\frac{\im \pi}{2\sinh(\pi p )}\left(I_{\im p}- \overline{I_{\im p}}\right).
\end{equation}
The Wronskian between the modified Bessel function of the first kind and its complex conjugate results to be
\begin{equation}
 W_z\left(I_{i|p|}(z),\overline{I_{i|p|}(z)}\right)=\frac{2\sinh(\pi p)}{i\pi z},
 \label{eq:Wr}
\end{equation}

Both these Bessel functions have an oscillatory behavior in a neighborhood of the origin $(\rho=0)$ \cite{AbSt:handbook},
\begin{align}
I_{\im p}(m\rho)  \approx \left( \frac{m\rho}{2} \right)^{\im p}/ \Gamma(\im p +1),\qquad 
K_{\im p}(m\rho)  \approx \sqrt{\frac{\pi}{p \sinh (\pi p)}} \cos \left( - p \ln \frac{m \rho}{2} + \arg \Gamma (\im p)\right).
\label{eq:asympIK0}
\end{align}
The behavior of the Bessel function $K_{\im p}$ for small value of the argument has been derived in \cite{SzBi:2009}.
For asymptotic values of their argument, the modified Bessel functions behave very differently,
\begin{equation}
I_{\im p}(m\rho) \approx \frac{e^{m\rho}}{\sqrt{2\pi m\rho}}, \qquad K_{\im p}(m\rho) \approx \sqrt{\frac{\pi}{2 m\rho}}e^{-m\rho}, \qquad \text{for $\rho \gg 1$}.
\label{eq:asympIK}
\end{equation}

The MacDonald function satisfies the identity
\begin{equation}\label{eq:mcdonnaldnorm}
\int_0^{\infty} \frac{\xd \rho}{\rho} \, K_{\im \mu}(\rho) \, K_{\im \mu'}(\rho) \, \frac{2 \mu \sinh (\mu \pi)}{\pi^2} = \delta(\mu - \mu'),
\end{equation}
which allows us to expand the field configuration $\varphi(\rho)$ on the hypersurface of constant Rindler time as
\begin{equation}
\varphi(\rho) = \int \xd p \, \varphi(p) \, \frac{\sqrt{2 p \, \sinh (p \pi)}}{\pi} \, K_{\im p} (m \rho), \qquad p \geq 0.
\label{eq:expansionK}
\end{equation}

\bibliographystyle{ieeetr}
\bibliography{hyperbola}
\end{document}